\documentclass{article}
\usepackage{jamaica04}
\usepackage{graphicx}
\usepackage{amssymb}
\usepackage{graphicx}
\frompage{000} \topage{000}                                              
\def\rightmark{Resonance Production in Heavy Ion Collisions}
\def\leftmark{C.Markert for the STAR collaboration}

\title{What do we learn from Resonance Production \\
in Heavy Ion Collisions ?}

\authors{
{Christina Markert$^1$ \dag\ for the STAR collaboration
}\\[2.812mm]
{\normalsize
\hspace*{-8pt}$^1$ Physics Department, Yale University, \\
New Haven, CT 06520,USA}}

\abstract{The resonance production of  $\Delta$, K(892),
$\Sigma$(1385), $\Lambda$(1520) and $\phi$ from elementary p+p and
Au+Au collisions at $\sqrt{s_{\rm NN}} = $ 200 GeV from the STAR
experiment at RHIC is presented. Yields and spectra are discussed
in terms of chemical and thermal freeze-out conditions. Thermal
models do not describe sufficiently the yields of the resonance
production in central Au+Au collisions. The approach which
includes pseudo-elastic and elastic hadronic interactions after
chemical freeze-out until thermal freeze-out suggests a time span
of $\Delta \tau>$5~fm/c.}

\PACS{specifications see, e.g.\ {\tt
http://www.aip.org/pacs/}}

\begin{document}

\maketitle
\setcounter{page}{1}

\section{Introduction}\label{intro}

Resonances have been proposed as a signature of a strongly
interacting nuclear medium due to modification of their properties
in the medium \cite{rap00}. The leptonic decay products of
resonances may provide a direct measurement of the hadronic stage
due to their very small final state interaction cross section.
However, hadronic decay products of resonances interact with
hadrons of the medium. Therefore they maybe useful in describing
the evolution of the source at this hadronic stage in terms of the
temperature and lifetime between chemical and thermal freeze-out.
The fact that in elementary p+p interactions the thermal
freeze-out temperature, from a statistical model, is close to the
chemical freeze-out temperature suggests that the life span is
close to zero. Comparison of the resonance yields and spectra from
Au+Au to p+p collision systems may indicate the influence of the
hadronic interacting medium. Microscopic model calculations give
suppression or enhancement factors for the resonance yields and
modifications of the momentum spectra due to rescattering and
regeneration of resonances in the hadronic medium
\cite{ble02,ble03}. The suppression is present because the
resonance invariant mass cannot be reconstructed by their decay
daughter if one of them rescatters in the medium.


\section{Data Analysis}

The $\Delta$, K(892), $\Sigma$(1385), $\Lambda$(1520) and $\phi$
resonances are reconstructed by measuring their decay daughters
($\Delta$$\rightarrow$p+$\pi$, K(892)$\rightarrow$K+$\pi$,
$\Lambda$(1520)$\rightarrow$ p+K and $\phi$$\rightarrow$K+K) with
the STAR {\em Time Projection Chamber} (TPC). Charged decay
particles are identified via energy loss ({\em dE/dx}) and their
measured momenta . The neutral strong decays such as $\Lambda$
from a $\Sigma$(1385) decay are reconstructed via a topological
analysis ($\Sigma$(1385) $\rightarrow$
$\Lambda$($\rightarrow$p+$\pi$) + $\pi$). The resonance signal is
obtained by the invariant mass reconstruction of each daughter
combination and subtraction of the combinatorial background
calculated by mixed event or like-sign techniques \cite{gau03}.
The resonance ratios, spectra and yields are measured at
mid-rapidity. The central trigger selection for Au+Au collisions
takes 5\% of the most central inelastic interactions. The setup
for the p+p interactions is a minimum bias trigger.

\section{Resonance yields in p+p and Au+Au collisions}

\begin{figure}[htb]
\centering
\includegraphics[width=0.8\textwidth]{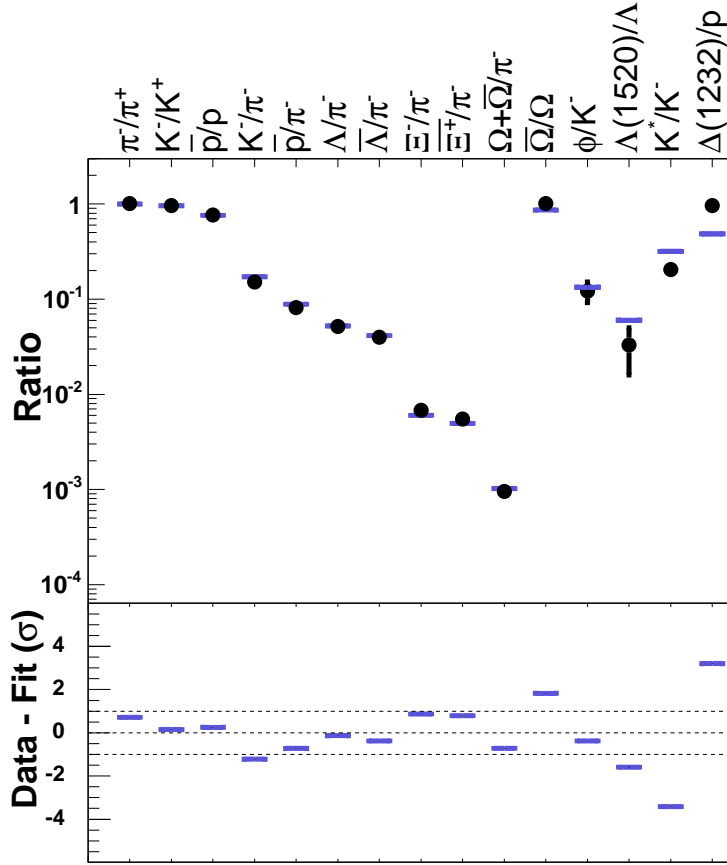}
 \caption{Particle ratios in central Au+Au collisions at
$\sqrt{s_{\rm NN}} = $ 200 GeV. The lines are the predictions of a
statistical model with T~=~160~$pm$~5, $\mu_{\rm B}$~=~24~$\pm$~4,
fitted to the stable particles for Au+Au collisions \cite{bra04}.}
 \label{thermal}
\end{figure}

The resonance and stable particle yields in p+p collisions at
$\sqrt{s_{\rm NN}} = $ 200 GeV are well described by a statistical
model with T=170 MeV \cite{bec02}. This is expected if the
particles are described by a statistical model in terms of phase
space and if there is no hadronic phase after chemical freeze-out.
The hadronic phase where the pseudo elastic and elastic
interactions (rescattering and regeneration) occur dominantly. The
statistical model fits the stable particles for $\sqrt{s_{\rm NN}}
= $ 200 GeV Au+Au with T=160-170 MeV ~\cite{bra04,pbm01,pbm03},
but the resonances are not well described by the model predictions
(see figure~\ref{thermal}).

\begin{figure}[htb]
 \centering
\includegraphics[width=0.6\textwidth]{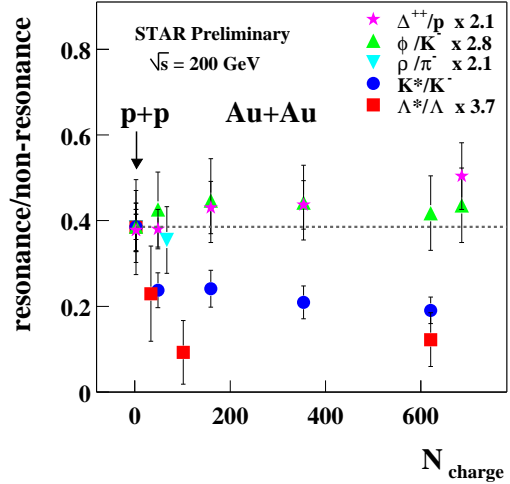}
 \caption{Resonance/non-resonance ratios of
$\phi$/K$^{-}$ \cite{ma03}, $\Delta^{++}$/p \cite{zha04},
$\rho/\pi$ \cite{fac03}, K(892)/K$^{-}$ \cite{zha03} and
$\Lambda$(1520)/$\Lambda$ \cite{gau03,mar03} for p+p and Au+Au
collisions at $\sqrt{s_{\rm NN}} = $ 200 GeV. The ratios are
normalized to the ratio of K*(892)/K$^{-}$ in p+p collisions.
Statistical and systematic errors are included.}
 \label{part}
\end{figure}

Microscopic model calculations (UrQMD \cite{ble02,ble03}) can
describe the decrease for the $\Lambda$(1520)/$\Lambda$ (UrQMD:
$\Lambda$(1520)/$\Lambda$=0.03) and the K(892)/K$^{-}$ (UrQMD:
K(892)/K$^{-}$ = 0.25), as they include rescattering of the decay
daughters. UrQMD predicts a slight increase for the
$\Delta^{++}$/p from p+p to Au+Au collision systems. This is in
agreement with the data (see figure~\ref{part}), which means
either the regeneration of the signal is of the same order as the
rescattering or there is no rescattering. The $\Delta^{++}$ has a
short lifetime of 1.3 fm/c, therefore the rescattering of the
decay daughters is expected to be large.
 Due to the long lifetime of the $\phi$ resonance (44 fm/c),
we would expect no significant signal loss by rescattering of the
daughters, since most of the decays happen outside the fireball
(10 fm/c), which is in agreement with our observation. The
measured values of K(892)/K and $\Lambda$(1520)/$\Lambda$ ratio
suggest a lifetime interval between chemical and kinetic
freeze-out greater than 5 fm/c when a thermal model with
rescattering but no regeneration is used \cite{tor01,mar02}.
\newpage

\vspace{0.5cm}

The decrease of the $\Lambda$(1520)/$\Lambda$ and K(892)/K$^{-}$
ratios from p+p to Au+Au collisions remains nearly constant from
peripheral collisions (50-80\%) up to the most central collisions
(5\%) (see figure~\ref{part}). This result suggests the same
$\Delta\tau$ for peripheral and central Au+Au collisions the same
lifetime of the system between chemical and thermal freeze-out.

\section{Resonance momentum spectra in p+p and Au+Au collisions}

\begin{figure}[htb]
\begin{minipage}[b]{0.5\linewidth}
\centering
\includegraphics[width=1.0\textwidth]{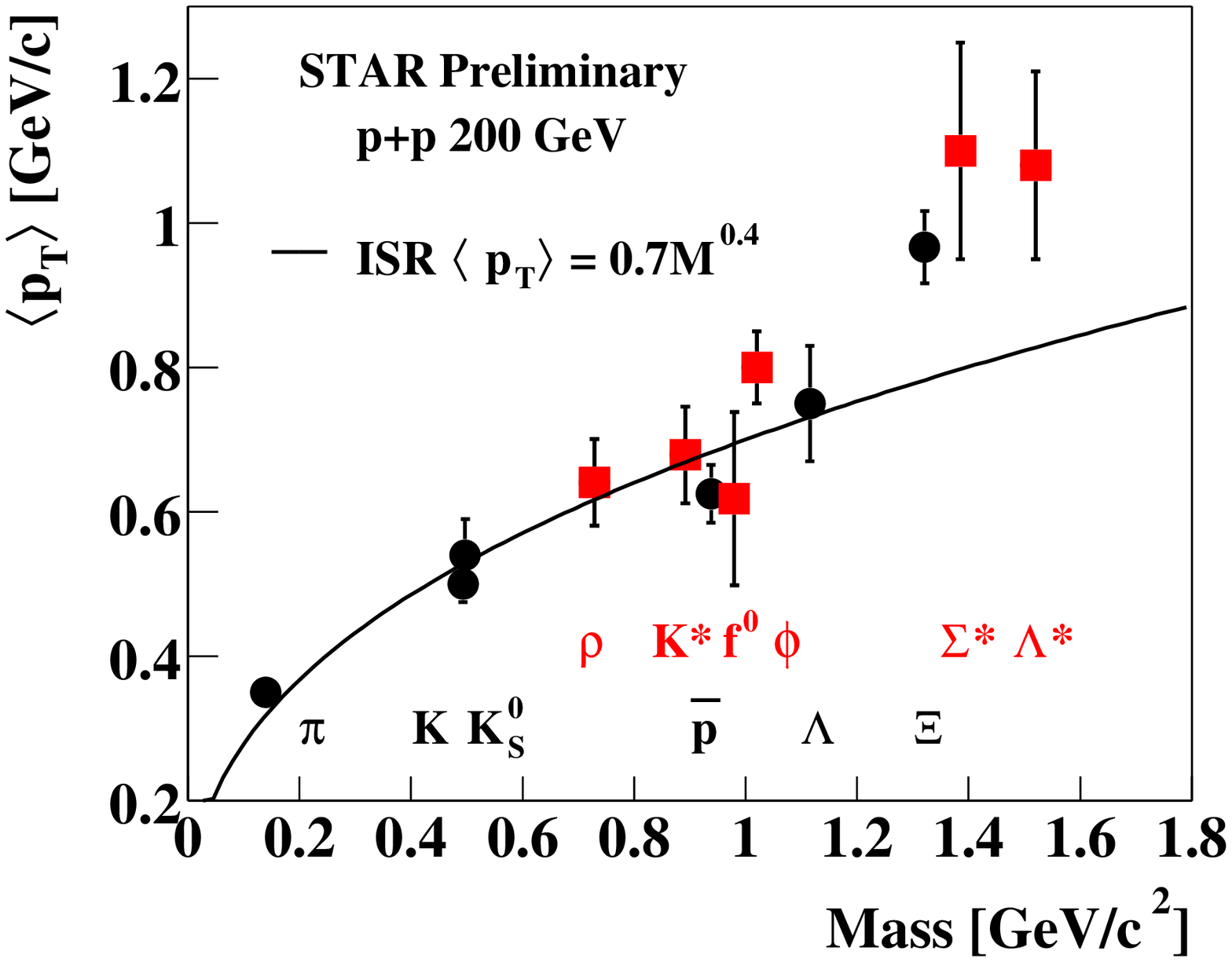}
\end{minipage}
\begin{minipage}[b]{0.5\linewidth}
\centering
\includegraphics[width=0.85\textwidth]{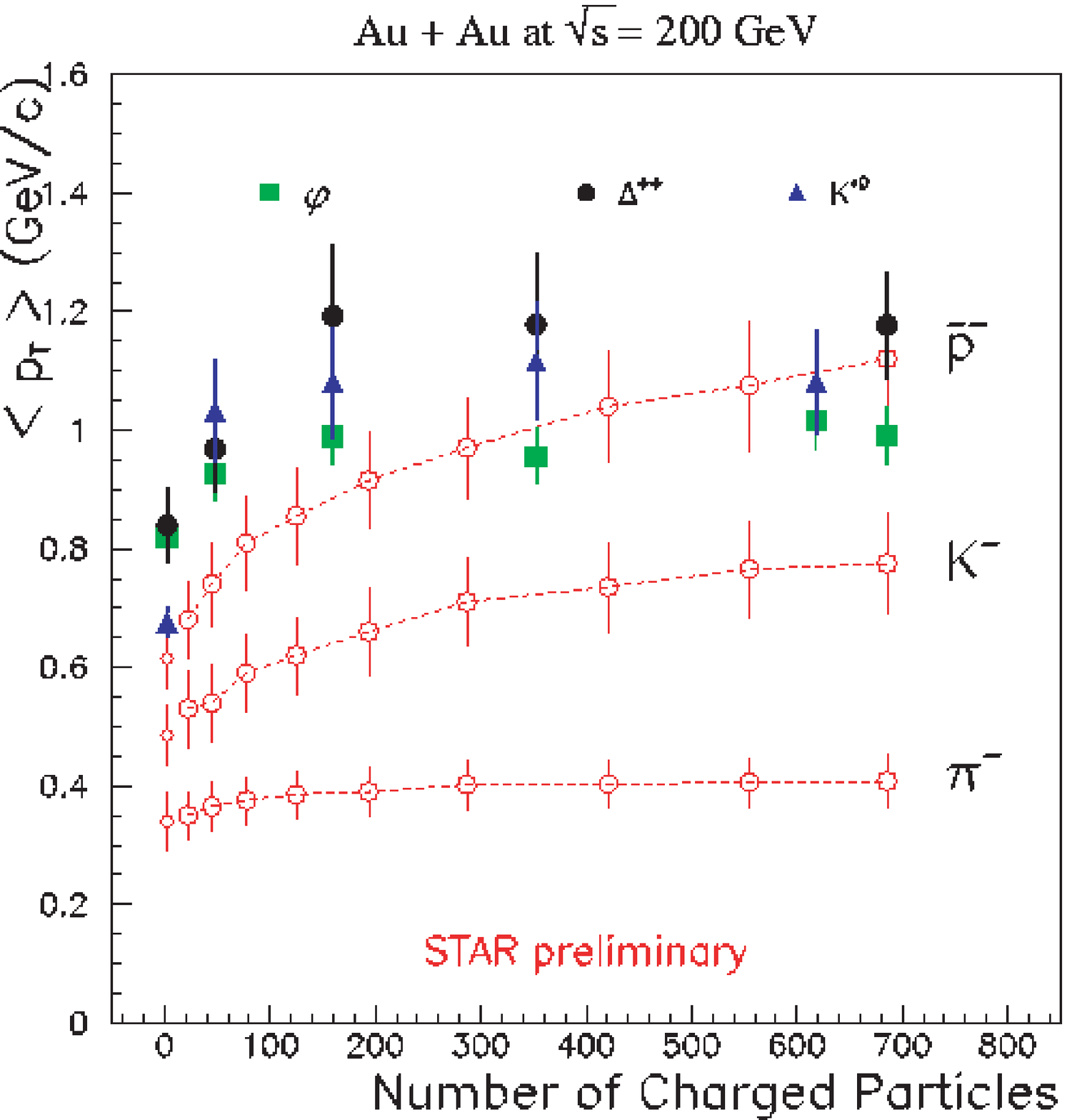}
\end{minipage}
\caption{Left: Mean transverse momentum, $\langle$p$_{\rm
T}$$\rangle$, as a function of particle mass in p+p collisions at
$\sqrt{s_{\rm NN}}=$200 GeV. The curve is a fit to
 measured ISR p+p data ($\pi$, K and p) at a collision energy of
 $\sqrt{s_{\rm NN}}=$26 GeV \cite{bou76}. Right: Mean transverse
 momentum $\langle$p$_{\rm T}$$\rangle$ as a function of charged
 particles for p+p and Au+Au collision systems for $\pi$, K, p, $\phi$, K(892) and $\Delta$.}
 \label{meanpt}
\end{figure}

The mean transverse momentum, $\langle$p$_{\rm T}$$\rangle$, as a
function of mass in figure~\ref{meanpt} ~\cite{mar04} (left panel)
shows a mass dependence for stable (circles) and resonance
(squares) particles. A fit to the ISR data shown by the curve
includes $\pi$, K and p from p+p collisions at $\sqrt{s_{\rm
NN}}=$26 GeV \cite{bou76} and is in good agreement with the $\pi$,
K and p data from $\sqrt{s_{\rm NN}}=$200 GeV. The resonances with
masses higher than 1 GeV/c$^{2}$ and the $\Xi$ indicate a stronger
$\langle$p$_{\rm T}$$\rangle$ dependence which is not represented
by the fit to the ISR data. The higher mass particles are more
dominantly produced in the higher multiplicity events which
introduces an increase in their $\langle$p$_{\rm T}$$\rangle$ and
a bias in the event selection.

According to microscopic model predictions the loss of signal in
the low momentum region would result in an increase of the inverse
slope parameter (and $\langle$p$_{\rm T}$$\rangle$) of the
transverse momentum spectra for the resonances. The
$\langle$p$_{\rm T}$$\rangle$ of the stable particles shows a
smooth rise from p+p to the most central Au+Au collisions. while
the resonances show a steep rise and stay constant up to the most
central Au+Au collision. This observation is in agreement with the
rescattering and regeneration of the resonance signal in the low
momentum region. The effect is much more pronounced for the K(892)
than for the $\phi$, due to the much shorter lifetime of K(892)
and the larger probability of signal loss due to rescattering.
This is also in agreement with the measured yields of the
resonances.

\vspace{0.5cm}

\section{Conclusions}\label{concl}

Resonances with their hadronic decay daughters are a unique tool
to probe the time span, $\Delta\tau$, between chemical and thermal
freeze-out in heavy ion collisions. The measured
$\Lambda$(1520)/$\Lambda$ and K(892)/K together with a thermal
model and rescattering suggest a lower limit of $\Delta\tau$ $>$5
fm/c. The effect of rescattering in the transverse momentum
distribution of the resonances can be observed as an increase of
their $\langle$p$_{\rm T}$$\rangle$ from p+p to most peripheral
Au+Au collisions. The mass dependence of $\langle$p$_{\rm
T}$$\rangle$ in p+p collisions including the high mass particles
indicates a linear trend, which is different from the ISR
parameterization.

\section*{Acknowledgement(s)}
At first I would like to thank the organizers who invited me to
such a well organized workshop at such a nice place. My special
thanks go to my theory friends J\"org Aichelin, Marcus Bleicher
and Horst St\"ocker. I would also like to thank Olga Barannikova,
Sevil Salur and Rene Bellwied for the many good and exiting
discussions that we had. And many thanks goes to the STAR
collaboration for the support in presenting this data.

\vfill\eject
\end{document}